\documentclass[12pt]{article}
\usepackage{amsfonts,epsfig}
\usepackage{amsmath,amsfonts}
\usepackage{amssymb,graphicx}

\def\pa{\partial}

\def\p{\varphi}

\def\d{\delta}

\def\b{\beta}

\def\f{\frac}

\def\p{\varphi}

\def\d{\delta}

\def\f{\frac}

\def\l{\label}
\def\e{\varepsilon}
\def\a{\alpha}

\def\m{\mu}
\def\n{\nu}

\def\r{\rho}
\def\s{\sigma}
\def\S{\Sigma}

\def\m{\mu}
\def\n{\nu}

\def\r{\rho}
\def\s{\sigma}
\def\S{\Sigma}

\def\o{\omega}
\def\sa{\stackrel{\leftrightarrow}}

\def\be{\begin{equation}}
\def\ee{\end{equation}}
\def\ba{\begin{eqnarray}}
\def\ea{\end{eqnarray}}

\begin{document}

\begin{center}
{\large \bf Density matrix of radiation of black hole with fluctuating horizon}
 \end{center}
\vspace*{1cm}
\centerline{\bf Mikhail Z. Iofa 
\footnote{ e-mail:iofa@theory.sinp.msu.ru}} 
\centerline{Skobeltsyn Institute of Nuclear
Physics}
\centerline{Moscow State University}
\centerline{Moscow 119991, Russia}

\begin{abstract}

The density matrix of Hawking radiation is calculated in the model of black hole with
fluctuating horizon. Quantum fluctuations smear the classical horizon of black hole
and  modify the density matrix of radiation producing
the off-diagonal elements. The off-diagonal elements may store information
of correlations between radiation and black hole. 
The smeared  density matrix was constructed by
convolution of the density matrix calculated 
with the instantaneous horizon with the Gaussian  
distribution over the instantaneous horizons.
The distribution has the extremum 
at the classical radius 
of the black hole and the width of order of the Planck length. 
Calculations were performed in the model of black hole formed 
by the thin collapsing shell which follows a trajectory which is a solution of the 
matching equations connecting the interior and exterior geometries.

\end{abstract}

\section{Introduction.}

From the time of Hawking's discovery that black holes radiate with the
black-body radiation, the problem of information stored in a black hole \cite{haw1}
attracted much attention.
Different ideas were discussed, in particular those of remnants \cite{gidd,pres,gidd1},
"fuzziness" of the black hole \cite{mathur,mathur1} and refs. therein, 
quantum hair \cite{coleman,strom,compere} and refs.therein., 
and smearing of horizon by quantum fluctuations \cite{ford,bru1,bru2,bru3}.
 The underlying
idea of the last approach is that small fluctuations of the background geometry
lead to corrections to the form of the density matrix of radiation. These
corrections are supposed to account for correlations between the black hole and radiation
and contain the imprint of information thrown into the black hole with the collapsing matter.

The idea that horizon of the black hole is not located at the
rigid position naturally follows from the observation that a black
hole as a quantum object is described by the wave functional over
geometries \cite{carlip,hadad,medv}. In particular, the sum over horizon areas
yields the black hole entropy.

In  papers \cite{bru2,bru3} the density matrix of black hole radiation was calculated in a
model with fluctuating horizon. Horizon fluctuations modify the Hawking density matrix
producing off-diagonal elements. Horizon fluctuations were taken into account by 
convolution the density matrix calculated with the instantaneous horizon 
radius $R$ 
with the black hole wave function which was taken in the Gaussian form 
$\psi (R) = N^{-1/2} e^{-(R - 2MG )^2 /2\s^2}$. 
Effectively the wave function introduces the smearing of
the classical horizon radius $\bar{R}=2MG$. The width of the
distribution, $\s$ , was taken of order the Plank lengths $l_p$
\cite{ford,bru2,bru3}.
In paper \cite{ford} it was stated that the "horizon 
fluctuations do not invalidate  the semiclassical
derivation of the Hawking effect until the black hole mass
approaches the Planck mass".

In this note we  reconsider calculation the density matrix of radiation emitted from the black
hole formed by the collapsing shell. The shell is supposed to follow the
infalling trajectory which is the exact solution to the matching equations connecting
the interior (Minkowski) and exterior (Schwarzschild) geometries of the space-time
\cite{brout,padm}.
In this setting one can trace  propagation of a ray (we consider only s-modes)
through the shell from
the past to the future infinity. For the rays propagating in the vicinity of the horizon
we obtain an exact formula connecting  $v_{in}$ at the past infinity and $u_{out}$
at the future infinity.

We obtain the expression for the "smeared" density matrix of Hawking radiation of the 
black hole with the horizon smeared by  fluctuations. In the limit 
$\s/MG \rightarrow 0$   the smeared density matrix turns to the Hawking density matrix.
The smeared density matrix is not diagonal and can be expressed as a sum of
the "classical part" and off-diagonal correction which is roughly of 
order $O(\s/MG)$ of the classical part. As a function of of frequencies $\o_{1,2}$
of emitted quanta the distribution is concentrated around $\o_1/\o_2 =1$ with the 
width of order $(\s/MG)\ln^{1/2}(MG/\s)$. 

The paper is constituted as follows. In Sect. 2 we review the geometry of 
the thin collapsing shell which follows a trajectory consisting of two phases. 
The trajectory is a solution of the matching equations connecting the 
internal and external geometries of the shell. 
We trace propagation of a light ray from the past to future infinity. 
In Sect.3 we introduce the wave function of the shell which saturates the uncertainty
relations. In Sect.4, we calculate the density
matrix of black hole radiation smeared by horizon fluctuations. Following the approach of
paper \cite{agu} calculation is performed by two methods: by the " $i\e$" prescription and
by using the normal-ordered two-point function. In Sect.5, using the exact 
expressions for the smeared radiation density matrix, 
we study the diagonal "classical" part of the density matrix and the 
off-diagonal elements.     

\section{Geometry of the thin collapsing shell}

In this section we introduce notations and review the geometry of space with
collapsing thin spherical shell \cite{brout,padm}. Outside of the shell the exterior geometry is
Schwarzschild space-time, the interior geometry is Minkowsky space-time.
In the Eddington-Finkelstein coordinates the metric of the exterior space-time is
\be
\l{1.1}
ds^2_{(ext)}=-\left(1-R/r \right)dv^2 +2dv dr +r^2d\Omega^2,\qquad r>R
\ee
where
$$
v = t + x(r), \qquad
u = t - x(r).
$$
$$
x(r) = r +R\ln \left(r/R -1\right) .$$ and
$$
v-u = 2x(r).
$$
The metric of the interior space-time is
\be
\l{1.2}
ds^2_{(int )}=-dV^2 +2dVdr +r^2 d\Omega^2
,\ee
where
$$
V=T+r, \qquad U=T-r
.$$
The light rays propagate along the cones $v, u =const$ in the exterior and along
$V,U =const$ in the interior regions.

Trajectory of the shell is $r=R_s (\tau )$, where $\tau$ is proper
time on the shell. The matching conditions of geometries on the
shell, at $r=R_s$, are
 \be 
\l{1.3}
 dV-dU=2dR_s,\qquad
dv-du=\f{2dR_s}{1-R/R_s }, \qquad dUdV= (1-R/R_s )dudv, 
,\ee 
where
the differentials are taken along the trajectory. From the
matching conditions follow the equations 
\ba
 \l{1.4}
2R'_s (1-U' )={U'}^2 - (1-R/R_s ),\\
\l{1.5}
2\dot{R}_s (1-\dot{V}) = -\dot{V}^2 +(1-R/R_s   )
.\ea
Here prime and dot denote derivatives over $u$ and $v$ along the trajectory.
\begin{center}
\begin{figure}
\begin{center}
\includegraphics [scale=.30]{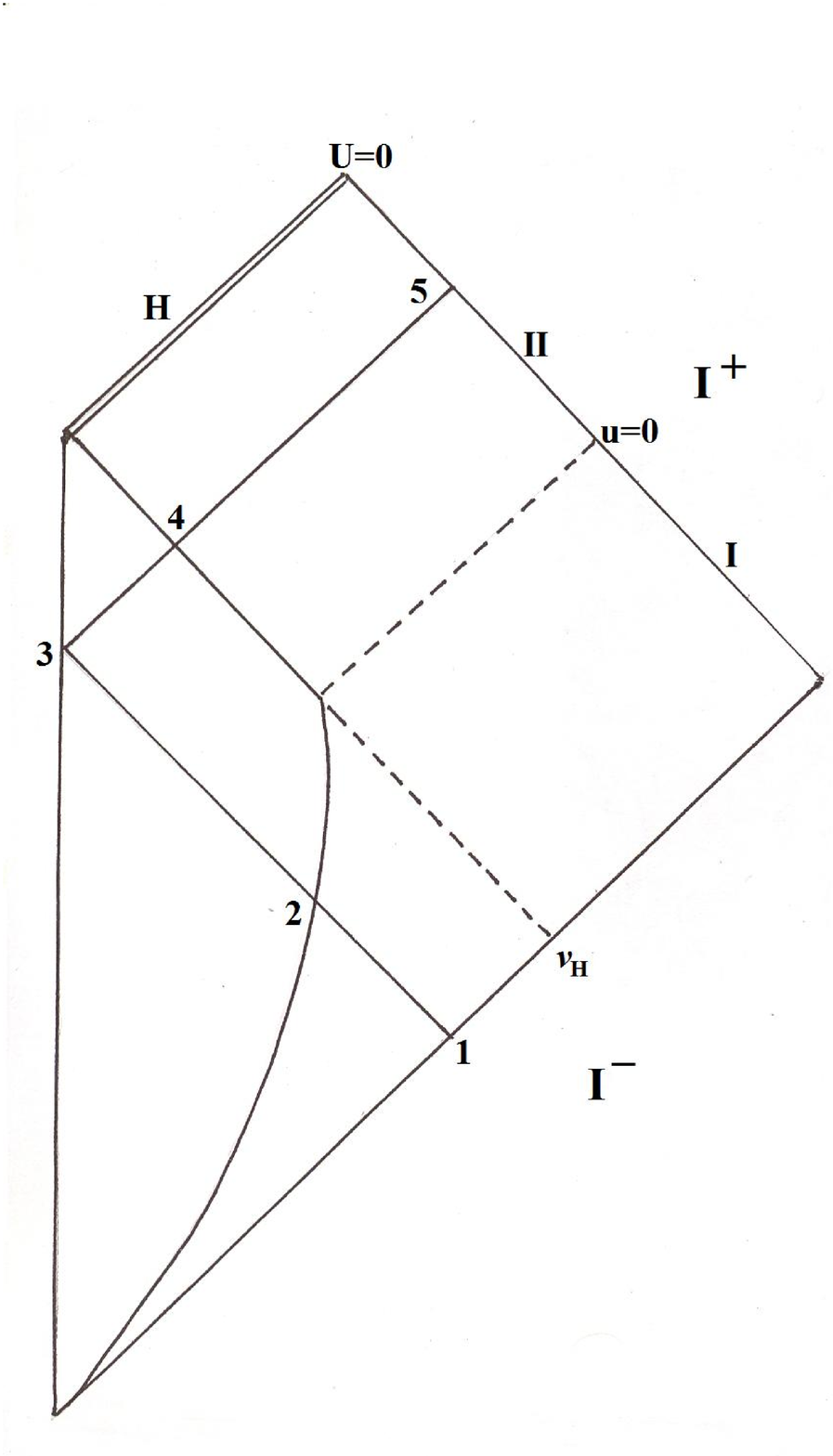}
\caption{Penrose diagram for collapsing shell.
For $u<0$ the shell is in the phase I,
for $u>0$ in the phase II. $v_H$ is the point of horizon formation.}
\end{center}
\end{figure}
\end{center}

The  trajectory of the shell consists of two phases \cite{padm}
$$
I. \,\quad u<0:\,\, R_s (u)=R_0 =const  .
$$
$$
II.\,\quad u>0: \,\, v=const,\,\, V=const.
$$
From the equations (\ref{1.4}),  (\ref{1.5}) are obtained the following expressions for
the trajectory;

In the phase I
\ba
\l{1.6}{}\,\,\,\,
{}\,\,\,\, U(u)=L_0 u -2R_0 +2R,\qquad  V(v)=L_0 (v-2x(R_0 ) )+2R,
\ea
where $L_0 =(1-R/R_0 )^{1/2}$.

In the phase II
\ba
\l{1.7}
{}\,\,\,\, V=2R, \qquad U= 2R-2R_s ,\\\nonumber
{}\,\,\,\, v=2x(R_0 ) \qquad u=2x(R_0 )-2x(R_s).
\ea
Horizon is formed at $U_H=0,\quad u\rightarrow \infty$ and $V_H =2R,\quad v_H =2x(R_0 )$.

We consider the modes propagating backwards in time. 
At $I^-$ the ray is in phase I,  after crossing the shell it
 reaches $I^+$  in the phase II.
Let the in-falling
 ray be at $I^-$ at $v_1<v_0$, where $v_0 =2x(R_0 )-2RL_0^{-1}$ 
is the point at which $V(v_0 )=0$
Between the points 1-2 the ray propagates outside the shell
in the phase I with $v=v_1$.
At the point 2 the ray crosses the shell and we have $v_2 =v_1 $ and
 $V_2 (v_1 ) =L_0 (v_1 -2x(R_0 ) )+2R = L_0 (v_1 -v_0 ) $.
The ray propagates in the interior of the shell, and at the point 3, at $r=0$, we have
$V_3 =V_2$. Reflection condition at the point 3 is $V_3 = U_3$.
At the crossing point 4 we have $ U_4 =U_3 $, where
$$
 U_4 =-2R_s (4) +2R,\qquad u_4 =-2x(R_s (4) )+2x(R_0 )
.$$
Here $R_s (4)$ stands for the radial position of the shell trajectory at the point 4.
The equation for $u_4$ can be written as
$$
\f{u_4}{2R}= \f{(R_0 -R)-(R_s -R)}{R} +\ln\f{R_0 -R}{R_s -R}
.$$
In the region $R_s (4) \sim R$, where $U_4\ll R$, 
neglecting in the first term $R_s -R$ as compared with $R_0 -R$, we obtain
the approximate equation for $u_4$ 
\be 
\l{1.8} 
\f{u_4}{2R}= \f{R_0 -R}{R} +\ln\f{-U_4/2}{R_0 -R}
\ee 
Thus, we have
\ba
v_1 -v_0 =L_0^{-1}V_2  =L_0^{-1}V_3=L_0^{-1}U_4 (u_4 ) \\\nonumber= 
-L_0^{-1}2(R_0 -R)e^{-(u_4 -2R_0)/R -1}
\l{1.8a}
\ea 
Removing the indices, we obtain our final result as
\be 
\l{1.10}
 v=v_0 -2(eL_0 )^{-1}(R_0 -R)e^{-(u -2R_0)/R} 
\ee
The above formulas are purely classical, modifications due to
back reaction of Hawking radiation are neglected.


\section{Quantum black hole}

Quantum nature of horizons of the black holes was discussed  in the
work of Carlip and Teitelboim \cite{carlip}, where it was shown
that the area of horizon $A$ and the opening angle, $\Theta$ or,
equivalently, the deficit angle $2\pi -\Theta$ form the canonical
pair. In paper \cite{bru3} it was shown that canonical
pair is formed by the opening angle and the Wald entropy $S_W$
\cite{wald}
\be
\l{4.1}
\left\{\Theta, \f{S_W}{2\pi}\right\} =1
.\ee

When the black hole is quantized, the Poisson bracket is
promoted to the commutation relation
\be
\l{4.2} 
[\hat{\Theta},\hat{S}_W ] =i\hbar .
\ee
The wave function of the black hole
satisfies the relation
\be
\l{4.3}
-i\f{\pa\Psi }{\pa S_W }=2\pi\hbar \Theta\Psi .
\ee
The minimal uncertainty $\Delta S_W \Delta \Theta =\hbar/2$ wave function is
\be
\Psi (\Theta )\sim e^{C(\Theta -2\pi )^2} e^{\f{i}{\hbar}} <S_W >\Theta ,
\ee
where $C\sim <S_W >$.
For the spherically symmetric configurations which
we consider the wave
function written through the instantaneous  horizon radius $R$ is
\be
\l{4.41} |\Psi (R)| = N^{-1}
e^{-\f{(R-\bar{R})^2}{4\s^2}}
.\ee
 The scale of
horizon fluctuations is
 $\s \sim l_p $ \cite{ford}, where $l^2_p =\hbar G$ is the
Planck length and $\bar{R} =2MG$ is the classical horizon radius
of the black hole of the mass $M$. The normalization factor $N$ is
\be
\l{4.5}
N^{-2}=\int ^\infty_0 4\pi dR R^2 e^{\f{(R
-\bar{R})^2}{2\s^2}} \simeq \s \bar{R}^2
.\ee

\section{Hawking radiation from the black hole formed by the shell}

Let us turn  the calculation  of Hawking radiation of the massless
real scalar field in the background of the black hole formed by
the shell. To perform quantization of the field, we restrict
ourselves to the $s$-wave modes. Expanding the scalar field in the
orthonormal set of solutions $u_i^-$ of the Klein-Gordon equation
which at the past null infinity $I^-$ have only positive frequency
modes we have 
\be 
\l{2.1} \p = \sum_i (a_i u^{(-)}_i + a^+_i u^{(-)*}_i ) .
\ee 
The scalar product of the fields is
\be
\l{2.1a}
(\p_1 ,\p_2 )=i\int_\Sigma d\Sigma^\m\p_2^* \sa\pa_\m \p_1
.\ee
Alternatively the field $\p$ can be expanded at
the hypersurface $\S^+ =I^+ \oplus H^+$ where $I^+$ is the future
null infinity and $H^+$ is the event horizon 
\be 
\l{2.2} 
\p = \sum_i (b_i u^{(+)}_i + b^+_i u^{(+)*}_i +c_i q_i + c^+_i q^*_i  ).
\ee 
Here $\{u^{(+)}_i\}$ is the orthonormal set of modes which
contain at the $I^+$ only positive frequencies and $\{q_i\}$ is
the orthonormal set of solutions of the wave equation which
contains no outgoing components \cite{haw1}. The operators $a_i ,\, a^+_i$ and
$b_i ,\, b^+_i$ are quantized with respect to the vacua $|in>$ and
$|out>$ correspondingly.

The modes $u^{(+)}_i$ can be expanded in terms of the modes $u^{(-)}_i$
\be
\l{2.3}
u^{(+)}_i =\sum_j (\a_{ij} u^{(-)}_j +\b_{ij} u^{(-)*}_j ),
\ee
where $\a_{ij}$ and $\b_{ij}$ are given by the scalar products
$$
\a_{ij}=(u_i^{(+)} , u^{(-)*}_j ), \qquad \b_{ij}=-(u_i^{(+)} , u^{(-)}_j ).
$$
For the spherically-symmetric collapse, the basis for the in- and outgoing modes is
$$
 u^{(-)}_{\o lm}|_{I^-} \sim \f{1}{\sqrt{4\pi\o}}\f{e^{-i\o v}}{r}Y_{lm}(\theta,\p ),
\qquad
 u^{(-)}_{\o lm}|_{I^+} \sim \f{1}{\sqrt{4\pi\o}}\f{e^{-i\o u}}{r}Y_{lm}(\theta,\p )
.$$
 Omitting the angular parts, the modes
$u^{(-)}_\o$ and $u^{(+)}_\o$ are
\ba
\l{2.4}
u^{(-)}_\o (v)|_{I^-} \sim \f{1}{\sqrt{4\pi\o}}{e^{-i\o v}},\qquad
u^{(+)}_\o (u)|_{I^+} \sim \f{1}{\sqrt{4\pi\o}}{e^{-i\o u}} . 
\ea
From (\ref{1.10}) we find
$$
u(v)=2R_0  +2R\left[-\ln (eL_0 ) +\ln \f{R_0 -R}{R} -\ln \f{v_0 -v }{2R}\right]=F(R)-2R\ln \f{R_0 -R}{R},
$$
where $F(R)= 2R_0  +2R[C +\ln ((R_0 -R)/{R})],\,\, C=-\ln (eL_0 )$. 
To simplify formulas, we consider the case
$R_0 \gg R$, so $\ln ((R_0 -R)/{R})\simeq \ln R_0 /R$, and
$$
F(R)\simeq 2R_0  +2R[C +\ln (R_0/R )
.$$
Note that both $v_0$ and $u(v)$ have explicit dependence on $R$.
 
The Bogolubov coefficient
 $$
\b_{\o_1\o_2} =i\int dv u^{(+)}_{\o_1}(v)\sa\pa_v u^{(-)}_{\o_2}(v) =
i\int \f{dv}{\sqrt{\o_1\o_2}}e^{-i\o_1 u(v)}\sa\pa_v e^{-i\o_2 v} 
 $$
 smeared by horizon fluctuations is obtained by convoluting it with
 the function $|\Psi^2|$
\footnote{Hereafter we abandon the numerical and greybody
factors.}
\ba 
\l{5.1} 
&{}&\bar{\b}_{\o_1\o_2}= \int_0^\infty dR R^2
e^{-(R -\bar{R} )^2 /2\s^2}N^{2} \b_{\o_1\o_2}\sim\\\nonumber 
&{}&\sim
(\o_2 /\o_1 )^{1/2}\int_{-\infty}^{v_0} dv
 e^{-i\o_1 (F(R) -2R \ln( (v_0 -v)/2R )-i\o_2 v}dR N^{2} R^2 e^{-(R -\bar{R} )^2/2\s^2} 
.\ea
Direct evaluation of the smeared Bogolubov coefficient (\ref{5.1}) yields (cf.\cite{agu})
\ba\l{5.1n}
\bar{\b}_{\o_1\o_2}\sim
\int dR R^2 N^2 e^{-(R -\bar{R} )^2 /2\s^2}R \sqrt{\f{\o_1}{\o_2}}
\f{e^{-i\o_1 F(R) +i\o_2 v_0} \Gamma (2Ri\o_1 )} {(-2Ri\o_2 +\e)^{2Ri\o_1}},
\ea

\subsection{Method 1}
Following the paper \cite{agu} we consider two ways of calculating the 
density matrix
\ba
\r_{\o_1\o_2} =i\int_{I^-} dv u^{(+)}_{\o_1} (v)\sa\pa_v u^{(+)*}_{\o_2} (v)=
i\int_{I^-} dv\int d\o' d\o'' \b_{\o_1\o'}\b^*_{\o_2 \o''}
u^{(-)}_{\o'} (v) \sa\pa_v u^{(-)*}_{\o''} (v)\\\nonumber
=
\int d\o' \b_{\o_1{\o'}}\b^*_{\o_2 {\o'}}\l{3.1}
,\ea
where in the last equality we used that the modes $\{u_\o^{(-)}\}$ form the
orthonormal set of functions on $I^-$.
The density matrix smeared by horizon fluctuations is
\be 
\l{m1}
\bar{\r}_{\o_1\o_2}=\int d{\o} ' \b_{\o_1{\o}' }\b^*_{\o_2 {\o}' }
\int_0^\infty dR_1 R_1^2 e^{-(R_1 -\bar{R} )^2 /2\s^2}N^{2}
\int_0^\infty dR_2 R_2^2 e^{-(R_2 -\bar{R} )^2
/2\s^2}N^{2} 
. \ee
Substituting (\ref{5.1n}), we have
\ba 
\l{m2}
\bar{\r}_{\o_1\o_2}\sim
\int_0^\infty
d\o'\f{\sqrt{\o_1\o_2}}{\o'}R_1 R_2  (-2iR_1\o' +\e)^{-2iR_1 \o_1}
(2iR_2\o'+\e)^{2iR_2\o_2} \Gamma(-2iR_1\o_1 ) \Gamma (2iR_2\o_2)
\\\nonumber
\times e^{-i\o_1 F(R_1 )  +i\o_2  F(R_2 )}dR_1 dR_2 R_1^2 R_2^2 N^4 e^{-(R_1
-\bar{R} )^2 /2\s^2} e^{-(R_2 -\bar{R} )^2
/2\s^2}
 \ea
The terms with $v_0$ have cancelled. Integrating over $\o'$, we obtain
\ba 
\l{m3}
\bar{\r}_{\o_1\o_2 }\sim\sqrt{\o_1\o_2}\int dR_1 dR_2 R^2_1 R^2_2\d (R_1\o_1 -R_2\o_2 )
\left(-\f{R_1}{R_2}\right)^{-2iR_1\o_1}\\\nonumber
\times \f{1}{R_1\o_1 \sinh ( 2\pi R_1\o_1 ) }
 e^{-i\o_1 F(R_1 )  +i\o_2  F(R_2 )}N^4 e^{-(R_1 -\bar{R} )^2
/2\s^2}e^{-(R_2 -\bar{R} )^2 /2\s^2}
\\\nonumber
=\sqrt{\o_1\o_2}\int dR_1 dR_2 R_1^3
R_2^3 \f{1}{R_1\o_1\o_2} \d \left(R_2 -R_1\f{\o_1}{\o_2} \right)
\\\nonumber\times
\left (e^{4\pi R_1 \o_1}-1 \right)^{-1}
e^{-2iR_0 (\o_1 -\o_2 )}N^4
e^{-(R_1 -\bar{R})^2/2\s^2 }e^{-(R_2 -\bar{R} )^2 /2\s^2 } 
,\ea
where, taking into account the $\d$-function, we substituted 
\ba
\nonumber
&{}&(-2iR_1 )^{-2iR_1\o_1}(2iR_2 )^{2iR_2\o_2}\Rightarrow
(-R_1/R_2 )^{-2iR_1\o_1}= e^{2\pi R_1\o_1}(R_1/R_2)^{-2iR_1\o_1}\\\nonumber
&{}&\Gamma(-2iR_1\o_1 ) \Gamma (2iR_2\o_2)\Rightarrow \f{1}{2R_1\o_1\sinh(2\pi R_1\o_1)}
\ea 
and 
\be\nonumber
e^{-i\o_1 F(R_1 )  +i\o_2  F(R_2 )}\Rightarrow 
e^{2iR_1\o_1\ln R_1 -2iR_2\o_2\ln R_2}=(R_1/R_2 )^{2iR_1\o_1}. 
\ee
Integration over $R_2$ yields 
\ba 
\l{m4}
&{}&\bar{\r}_{\o_1\o_2}\sim\f{1}{\sqrt{\o_1\o_2}}\int
dR_1 R_1^5 \left(\f{\o_1}{\o_2}\right)^{3}
\left( e^{4\pi R_1\o_1 }-1 \right)^{-1}
e^{ -2iR_0 (\o_1 -\o_2 ) }\\\nonumber
&{}&\times \f{1}{\bar{R}^4 \s^2}\exp{\left\{-\f{1}{2\s^2}\left(R_1 \sqrt{1+{\o_1^2}/{\o_2^2}}-
\bar{R}\f{1+\o_1/\o_2}{\sqrt{1+\o_1^2/\o_2^2 }}\right)^2 \right\}}
\exp{\left\{- \f{\bar{R}^2 (\o_1 -\o_2 )^2 }{ 2\s^2 (\o_1^2 +\o_2^2 )}\right\}} 
\ea
Because $\bar{R}/\s \gg 1$ both exponents have sharp extrema. Integrating over $R_1$,
we arrive to the density matrix of the form
\ba
\l{m5}
&{}&\bar{\r}_{\o_1\o_2}\sim\f{1}{\sqrt{\o_1\o_2}}\left(\f{\o_1}{\o_2}\right)^{3}
\left(\f{1+\o_1/\o_2}{1+\o_1^2/\o_2^2}\right)^5 \f{\bar{R}}{\s(1+\o_1^2/\o_2^2 )^{1/2}}
\left(e^{4\pi \bar{R} (\o_1 +\o_2 )\o_1\o_2/(\o_1^2 +\o_2^2 )} -1\right)^{-1}
\\\nonumber
&{}&\times
e^{- 2iR_0  (\o_1 -\o_2 ) }
\exp{\left\{- \f{\bar{R}^2 (\o_1 -\o_2 )^2 }{ 2\s^2 (\o_1^2 +\o_2^2 )}\right\}}
\ea


\subsection{Method 2}

Alternatively, the density matrix can be presented in the
following form 
\be \l{3.2} 
\bar{\r}_{\o\o'}=\int d\o_1 \int_\S d\S_1^\m
u^{(+)*}_{\o lm}(x_1 )\sa\pa_\m u^{(-)}_{\o_1 l_1 m_1} (x_1 ) \int_\S d\S_2^\n
u^{(+)}_{\o' l' m'} (x_2 )\sa\pa_\n u^{(-)*}_{\o_1 l_1 m_1} (x_2 ) 
\ee
 where for
the initial value hypersurface can be taken either $I^-$ or $I^+$.
Expanding $\p$ in the basis $\{u^{(-)}\}$
$$
<in|\p (x_1 )\p (x_2 )|in>=\int d\o_1 u^{(-)}_{\o_1} (x_1 )u^{(-)*}_{\o_1} (x_2  )
$$
where $|in>$ and $|out>$ are vacuum states at $I^-$ and $I^+$,
and using the relation
$$
<in|:\p (x_1 )\p (x_2 ):|in>=<in|\p (x_1 )\p (x_2 )|in> -<out|\p (x_1 )\p (x_2 )|out>,
$$
we obtain
\be 
\l{3.3} 
\r_{\o_1\o_2}=\int_\S d\S_1^\m \int_\S
d\S_2^\n [u^{(+)*}_{\o_1} (x_1 )\sa\pa_\m ] [u^{(+)}_{\o_2} (x_2
)\sa\pa_\n ]<in|:\p (x_1 )\p (x_2 ):|in> 
\ee
 
To perform calculation of (\ref{3.3}) one can use the expansion of
the two-point function \newline $<in|:\p (x_1 )\p (x_2 ):|in>$ on
$I^+$ to obtain \cite{agu}
\ba 
\l{3.4}
 \r_{\o_1\o_2}\sim (\o_1 \o_2 )^{-1/2}\int_{I^+}du_1
du_2 e^{-iu_1\o_1 +iu_2\o_2} \left(\f{(dv/du) (u_1 )(dv/du) (u_2
)} {(v(u_1 )- v(u_2 )-i\e )^2}-\f{1}{(u_1 -u_2 -i\e )^2} \right)
.\ea
where for $v(u)$ we take the function (\ref{1.10}). 
For the density matrix modified by  horizon fluctuations we obtain
\ba 
\l{5.3}
&{}&\bar{\r}_{\o_1\o_2}\sim (\o_1 \o_2 )^{-1/2}
\int_{I^+}du_1\int_{I^+}
du_2 \f{e^{-iu_1\o_1 +iu_2\o_2}}{R_1 R_2} \f{ e^{- (u_1 -2R_0 ) /2R_1}  e^{-(u_2 -2R_0 ) /2R_2}} 
{\left( e^{- (u_1 -2R_0 ) /2R_1}  -
 e^{- (u_2 -2R_0 ) /2R_2} -i\e  \right)^2}
\\\nonumber
&{}&\times  e^{-(R_1 -\bar{R})^2/2\s^2 - (R_2 -\bar{R} )^2 /2\s^2 }N^4 R_1^2 R_2^2{d R_1}{dR_2}
,\ea
where for $v(u)$ is taken the function (\ref{1.10}).
Extracting in the denominator the
factor \newline $( e^{-(u_1-2R_0)/ 4R_1 -(u_2-2R_0 )/4R_2 })^2$, shifting 
$  u_i-2R_0 \Rightarrow u_i$ and changing variables
$u_i /4R_i \Rightarrow u_i$, we obtain
\ba 
\l{5.4}
 \bar{\r}_{\o_1\o_2}\sim (\o_1 \o_2 )^{-1/2}
\int^{\infty}_{-\infty} du_1\int^{\infty}_{-\infty} du_2
e^{-4i\o_1 u_1 R_1 +4i\o_2 u_2 R_2 -2iR_0 (\o_1 -\o_2 )}
\sinh^{-2}\left(u_1 -u_2  -i\e\right)
\\\nonumber 
\times e^{-(R_1 -\bar{R})^2/2\s^2 - (R_2 -\bar{R} )^2/2\s^2 }
N^{4}R_1^2 R_2^2 d R_1 dR_2 
\ea
Performing the contour integration over $u_1$ around the pole in
the upper half plane using the formula
$$
\int_{-\infty}^\infty dy \f{e^{-i\o y}}{\sinh^2 (y-z -i\e )}=2\pi\f{e^{-i\o z}}{e^{\pi\,\o} -1}
,$$
 we have
\ba
\l{5.5}
\bar{\r}_{\o_1\o_2}\sim (\o_1 \o_2 )^{-1/2}\int dR_1 dR_2 R_1^2 R_2^2 N^{4}
\o_1 R_1 \f{1}{e^{4\pi\o_1 R_1} -1} \int du_2
e^{-4i\o_1 R_1 u_2 +4i\o_2 R_2 u_2 }
\\\nonumber 
\times  e^{-2iR_0 (\o_1 -\o_2 )}  
e^{-(R_1 -\bar{R})^2/2\s^2 - (R_2 -\bar{R} )^2 /2\s^2 }.
\ea
 Integration over $u_2$ yields
 \ba
\l{5.6}
\bar{\r}_{\o_1\o_2}\sim (\o_1\o_2 )^{-1/2}  \int dR_1 dR_2 R_1^2 R_2^2 N^{4}
\o_1 R_1\f{1}{e^{4\pi\o_1 R_1} -1} \d (\o_1 R_1 -\o_2 R_2 ) 
\\\nonumber
\times e^{-2iR_0 (\o_1 -\o_2 )}
 e^{-(R_1 -\bar{R})^2/2\s^2 - (R_2 -\bar{R} )^2 )/2\s^2 } .
\ea
Integrating over $R_2$ and removing the $\d$-function, we obtain 
\ba 
\l{5.7}
\bar{\r}_{\o_1\o_2}\sim\f{1}{\sqrt{\o_1\o_2}}
\int dR_1 R_1^5\left(\f{\o_1}{\o_2}\right)^{3}
\left( e^{4\pi R_1\o_1}-1\right)^{-1}
e^{-2iR_0 (\o_1 -\o_2 )}\\\nonumber
\times\f{1}{\bar{R}^2 \s^2}\exp{ \left\{-\f{1}{2\s^2}\left( R_1\sqrt{1+{\o_1^2}/{\o_2^2}}-
\bar{R}\f{ 1+\o_1/\o_2 }{\sqrt{1+\o_1^2/\o_2^2 }}\right)^2\right\} }
\exp{\left\{-\f{\bar{R}^2 (\o_1 -\o_2 )^2}{2\s^2 (\o_1^2 +\o_2^2 )}\right\}} 
.\ea
Expression (\ref{5.7}) is identical to that obtained by method 1 in (\ref{m4}).

\section{Diagonal and off-diagonal parts of the density matrix}

In the limit $\bar{R}/\s \rightarrow \infty$ the expression 
\be
\l{5.71} 
\f{ \bar{R}}{\s \sqrt{1+\o_1^2/\o_2^2}}
\exp{\left\{-\f{\bar{R}^2}{2\s^2}\f{(1-\o_1/\o_2)^2}{1+\o_1^2/\o_2^2}\right\}} 
\ee
becomes  the delta function $\d (1-\o_1/\o_2 )$. 
The density matrix  $\bar{\r}_{\o_1\o_2}$ Eq. (\ref{5.7}) 
 turns into the formula for the Hawking spectrum
 \be 
\l{5.8}
\r_{\o_1\o_2} \sim \sqrt{ \f{\o_1}{\o_2 } }\d (\o_1 -\o_2 )
\left( e^{2\pi \bar{R} (\o_1 +\o_2 )}-1\right)^{-1} .
\ee

The smeared density matrix contains the off-diagonal elements. Because
the density matrix has the sharp maximum at $\o_1/\o_2 =1$, it is natural
to divide it into the "classical" contribution
\be 
\l{5.9}
\r_{\o_1\o_2}^{cl} \sim
\Theta\left(2\f{\s}{\bar{R}}\ln^{1/2}\f{\bar{R}}{\s}- \left|1-\f{\o_1}{\o_2} \right|\right) 
\f{\bar{R} }{\s} 
\exp\left\{-\f{\bar{R}^2 (1-\o_1/\o_2 )^2}{4\s^2}\right\} 
\sqrt{\f{\o_1}{\o_2}}\left( e^{2\pi \bar{R}(\o_1 +\o_2 )}-1 \right)^{-1} 
\ee 
and the off-diagonal correction.
As mentioned above,
in the classical contribution 
the factor
multiplying $\left(e^{2\pi \bar{R} (\o_1 +\o_2 )}-1\right)^{-1}$  in the limit $\bar{R}/\s
\rightarrow \infty$ turns into the  $\d$-function.

At $ \o_1/\o_2 =1$ the expression (\ref{5.71}) equals $\bar{R}/(\s\sqrt{2})$.
At $\o_1/\o_2 =1\pm 2(\s/\bar{R})\ln^{1/2}(\bar{R}/\s ) $ (\ref{5.71}) is of order unity.
Stated differently, at the distance $2(\s/\bar{R})\ln^{1/2}(\bar{R}/\s)$ from the 
extremum, the off-diagonal part is of order $O(\s/\bar{R})$ of the classical 
expression at the point of extremum.
 To make this
difference explicit we extract the factor ${\s}/{\bar{R}}$:
\be
 \l{5.11} 
\r_{\o_1 \o_2} = \r_{\o_1\o_2}^{cl}+\f{\s}{\bar{R}}\Delta \r_{\o_1\o_2} 
,\ee
where
 \be \l{5.10}
\f{\s}{\bar{R}}\Delta\r_{\o_1\o_2}\sim\r_{\o_1\o_2}
\Theta\left(\left|1-\f{\o_1}{\o_2}\right |-2\f{\s}{\bar{R}}\ln^{1/2}\f{\bar{R}}{\s}
\right). 
\ee 
It is of interest to evaluate the contribution of small distances
to  the smeared density matrix
(cf. \cite{agu}). It is convenient to use the method 2. Starting
from (\ref{5.3}) and making the change of variables $u_i \rightarrow
u_i R_i/\bar{R}$, we have
\ba 
\l{5.12}
\bar{\r}_{\o_1\o_2}\sim (\o_1\o_2 )^{-1/2} \int du_1\int du_2
e^{-4i\o_1 u_1 R_1/\bar{R} +4i\o_2 u_2 R_2/\bar{R}  }
{\sinh^{-2}\left(\f{(u_1 -u_2)}{\bar{R}}-i\e\right)}
\\\nonumber
\times N^{4}e^{-(R_1 -\bar{R})^2/2\s^2 - (R_2 -\bar{R} )^2/2\s^2 }
R_1^2 R_2^2 d R_1 dR_2
,\ea
 where we omitted the irrelevant for the estimate terms.

Introducing $z=(u_1 -u_2 )/2,\,\,y =(u_1 +u_2 )/2$, we integrate
over $y$ in the interval $(-\infty,\infty)$ and over $z$ in the
interval $(-\a,\a )$:
\ba
\l{5.13}
 \bar{\r}_{\o_1\o_2}\sim (\o_1 \o_2 )^{-1/2} \int_{-\a}^\a dz \f{R_1 R_2}{\bar{R}^2}
e^{-iz( R_1 \o_1 +R_2 \o_2 )/\bar{R} }\bar{R}
 \d (R_1\o_1 -R_2 \o_2 )\f{1}{\sinh^2 (z/R -i\e )}
\\\nonumber
\times N^{4}e^{-(R_1 -\bar{R})^2/2\s^2 - (R_2 -\bar{R} )^2/2\s^2 }
R_1^2 R_2^2 d R_1 dR_2 
\ea

Integrating over $R_2$, we obtain the expression structurally
similar to (\ref{m4}) and (\ref{5.7}). Because this expression has
sharp extremum at $R_1 =\bar{R}$ and $\o_1/\o_2 =1$, for our estimates we
can set in the integrand  $R_1$ and $\o_1/\o_2$ equal to
the extremal values.

The resulting density matrix is
\be 
\l{5.14}
 I\sim (\o_1\o_2 )^{-1/2}\int_{-\a}^\a dz e^{-i\o z}\f{\bar{R}^2}{\sinh^2 (2\bar{R}z )} \Theta
\left(\f{\s}{\bar{R}}\ln\f{\bar{R}}{\s} -\left|1 -\f{\o_1}{\o_2}\right |\right).
\ee
The integral in (\ref{5.14}) was estimated in \cite{agu} for $\o\bar{R}<1$ and
it was shown that that the ratio of (\ref{5.14}) to the Hawking spectrum is
\be 
\l{5.15} 
\f{I(\o \bar{R}, \a/\bar{R})}{\left(e^{4\pi\o\bar{R}} -1\right)^{-1}}
\sim \a/\bar{R} .
\ee

Taking $\a\sim \s\ln\bar{R}/\s$ and assuming for an estimate that the mass of the
black hole is of order of several solar masses, we obtain that 
$\a/\bar{R}\sim (\s/M)\ln(M/\s)\ll 1$.

\section{Discussion}

In this paper we discussed modifications of the density matrix of
radiation of the black hole formed by the collapsing shell
resulting from horizon fluctuations of black hole. Horizon
fluctuations are inherent to the black hole considered as a
quantum object. In distinction with the original Hawking
calculation based on the rigid horizon, horizon fluctuations
provide the off-diagonal matrix elements of the density matrix.
Qualitatively, the off-diagonal matrix elements account for
correlations between the particles in radiation and for
information stored in these correlations.

The  construction of the density matrix discussed in the present note
 is parallel to that of papers \cite{bru2,bru3},
where the density matrix with the off-diagonal corrections was obtained in the form
$\r_{H} (\o,\tilde{\o}) + C_{BH}^{1/2}\Delta\r (\o,\tilde{\o})$,
where $\r_{H}$
is the original Hawking matrix   and the off-diagonal correction is of order $C^{1/2}$, where
$C_{BH}=l_p^2 /4\pi M^2 G^2$, where $M$ is the mass of the shell.
The fact that the expansion parameter in both
approaches is the same is rather obvious because $\s/M$ is the
only dimensionless parameter connecting the horizon radius  and
the scale of fluctuations.

The  details of calculations and the actual form of the off-diagonal terms in
the density matrix obtained  in the present paper and in the
papers \cite{bru2} are different.

Because the structural form (but not the explicit form ) of the
smeared density matrix obtained in the present paper is similar to that
  in papers \cite{bru2,bru3}, we arrive at the same qualitative
conclusions concerning the information problem as in these papers. 
It is possible to construct the
$N$-particle density matrix $\r^{(N)}$ having dimensionality
$N\times N$ and to calculate the entropy of radiation $S/N =
-Tr(\r^{(N)}\ln\r^{(N)}$. Calculating the information contained in
radiation, which is defined as the difference between the thermal
Bekenstein-Hawking entropy $S_H$ of radiation,  $I= S_H -S$,
 one  obtains the qualitatively correct
 Page purification curve \cite{page1,page,harl}.

However, the above results pose a question. In \cite{gidd1,mathur} it was shown that
the Schwarzschield metric admits construction of "nice slices".The nice slices 
are at $r\simeq const$ inside the horizon, and one can take $r\sim M/2$. For 
$M\gg l_p$ horizon fluctuations which are on the scale $l_p$ are insignificant 
for particle production on the nice slices. If, however, the horizon fluctuations 
are somehow connected with hair (in spirit of \cite{strom,compere} and refs. therein), 
then the niceness is broken and horizon fluctuations can be connected with the release
of information from the black hole.   

The expressions for the density matrix  discussed
in the present paper refer to ethernal black holes. Because of the outgoing 
flux of particles, the mass of collapsing shell is not constant, but decreases
with time
$$
\f{\pa M(u)}{\pa t}=-<T_{uu}>\equiv -L_H.
$$
Here $T_{uu}$ is the $uu$ component of the radiation stress tensor.
In papers \cite{brout,padm} it was found that for the mass of black hole 
$M(u)\gg m_p$, where $m_p$ is the Planck mass, the backreaction of black hole radiation does 
not prevent formation of the event horizon.
When the outgoing flux is small and slowly varying, the calculation is self-consistent.
The metric of the exterior geometry of	the shell at large distances $r$ becomes
$$
ds^2\simeq -\left(1-\f{2M(u)}{r}\right)dv^2 +2du dr +r^2 d\Omega^2
$$
where
$dM(u)/du =-L_H$, and $L_H \sim 1/M^2 (u)$. 

For the case considered in Sect.2 at the leading order 
$$
L_H =\f{1}{48\pi}\f{M}{R_s^3}\left(2 -\f{3M}{R_s}\right),
$$
where $M$ is the mass of the shell. 
Substituting $R_s (U) =-(U-4M )/2$, we have
$$
L_H =\f{M}{3\pi}\f{U-M}{(U-4M)^4}.
$$
In the near-horizon region $U\rightarrow 0$, and we obtain $L_H \simeq 1/(768\pi M^2 )$.
This shows that our semiclassical treatment is valid.

\vspace*{0.5cm}

{\large\bf Acknoledgments}

This research was supported by the
Institute of Nuclear Physics of Moscow State University.

\end{document}